\documentclass[aps,prb,twocolumn,superscriptaddress,showpacs]{revtex4}
\usepackage{graphicx}
\usepackage{latexsym}
\usepackage{amssymb}
\usepackage{amsmath}
\usepackage{amsfonts}
\usepackage{bm}
\usepackage{multirow}
\usepackage{color}
\usepackage{comment}
\newcommand{\ii}{\mathrm{i}}

\newcommand{\SO}{\mathrm{SO}}
\renewcommand{\O}{\mathrm{O}}
\newcommand{\SU}{\mathrm{SU}}
\newcommand{\U}{\mathrm{U}}

\newcommand{\vect}[1]{{\bm{#1}}}

\newcommand{\figref}[1]{Fig.\,\ref{#1}}

\newcommand{\beq}{\begin{equation}}
\newcommand{\eeq}{\end{equation}}
\newcommand{\beqn}{\begin{eqnarray}}
\newcommand{\eeqn}{\end{eqnarray}}

\DeclareMathAlphabet{\mathbbold}{U}{bbold}{m}{n}

\def\SU{{\rm SU}}

\def\U{{\rm U}}

\begin{document}

\title{Topological Edge and Interface states at Bulk disorder-to-order \\ Quantum Critical Points}

\author{Yichen Xu}
\affiliation{Department of Physics, University of California,
Santa Barbara, CA 93106, USA}

\author{Xiao-Chuan Wu}
\affiliation{Department of Physics, University of California,
Santa Barbara, CA 93106, USA}

\author{Chao-Ming Jian}
\affiliation{Station Q, Microsoft, Santa Barbara, California
93106-6105, USA}

\author{Cenke Xu}
\affiliation{Department of Physics, University of California,
Santa Barbara, CA 93106, USA}

\begin{abstract}

We study the interplay between two nontrivial boundary effects:
(1) the two dimensional ($2d$) edge states of three dimensional
($3d$) strongly interacting bosonic symmetry protected topological
states, and (2) the boundary fluctuations of $3d$ bulk
disorder-to-order phase transitions. We then generalize our study
to $2d$ gapless states localized at an interface embedded in a
$3d$ bulk, when the bulk undergoes a quantum phase transition. Our
study is based on generic long wavelength descriptions of these
systems and controlled analytic calculations. Our results are
summarized as follows: ($i.$) The edge state of a prototype
bosonic symmetry protected states can be driven to a new fixed
point by coupling to the boundary fluctuations of a bulk quantum
phase transition; ($ii.$) the states localized at a $2d$ interface
of a $3d$ $\SU(N)$ quantum antiferromagnet may be driven to a new
fixed point by coupling to the bulk quantum critical modes.
Properties of the new fixed points identified are also studied.

\end{abstract}

\maketitle

\section{Introduction}

The most prominent feature of topological insulators
(TI)~\cite{kane2005a,kane2005b,fukane,moorebalents2007,ludwigclass1,ludwigclass2,kitaevclass}
and more generally symmetry protected topological (SPT)
states~\cite{wenspt,wenspt2} is the contrast between the boundary
and the bulk of the system. In particular the $2d$ edge of $3d$
SPT states hosts the most diverse zoo of exotic phenomena that
keep attracting attentions and efforts from theoretical physics.
It has been shown that many exotic phenomena such as anomalous
topological
order~\cite{TI_fidkowski1,TI_fidkowski2,TI_qi,TI_senthil,TI_max,wangsenthil,chengedge},
deconfined quantum critical points~\cite{senthilashvin}, self-dual
field theories~\cite{xudual,mrossdual,seiberg2,deconfinedual} can
all occur on the $2d$ edge of $3d$ SPT stats. Sometimes the
symmetry of the system is secretly realized as a self-dual
transformation of the field theories at the
boundary~\cite{seiberg1,dualreview}. All these suggest that the
$2d$ boundary of a $3d$ system is an ideal platform of studying
physics beyond the standard frameworks of condensed matter theory.

On the other hand, even the boundary of an ordinary
Landau-Ginzburg type of quantum phase transition can have
nontrivial behaviors. It was studied and understood in the past
that the boundary of a bulk conformal field theory (CFT) follows a
very different critical behavior from the
bulk~\cite{cardybook,cardyboundary,boundary2,boundary3,boundary4,boundary5},
due to the strong boundary condition imposed on the CFT. The
boundary fluctuations (or the boundary CFT) of the Landau-Ginzburg
phase transitions were studied through the standard
$\epsilon-$expansion, and it was shown that the critical exponents
are very different from the bulk. Hence if experiments are
performed at the boundary of the system, one should refer to the
predictions of the boundary instead of the bulk CFT. These two
different boundary effects were studied separately in the past. In
this work we will study the interplay of these two
distinct boundary effects. 
Our goal is to seek for new physics, ideally new fixed points
under renormalization group (RG) flow due to the coupling of the
two boundary effects.

\begin{figure}[h]
\includegraphics[width=0.95\linewidth]{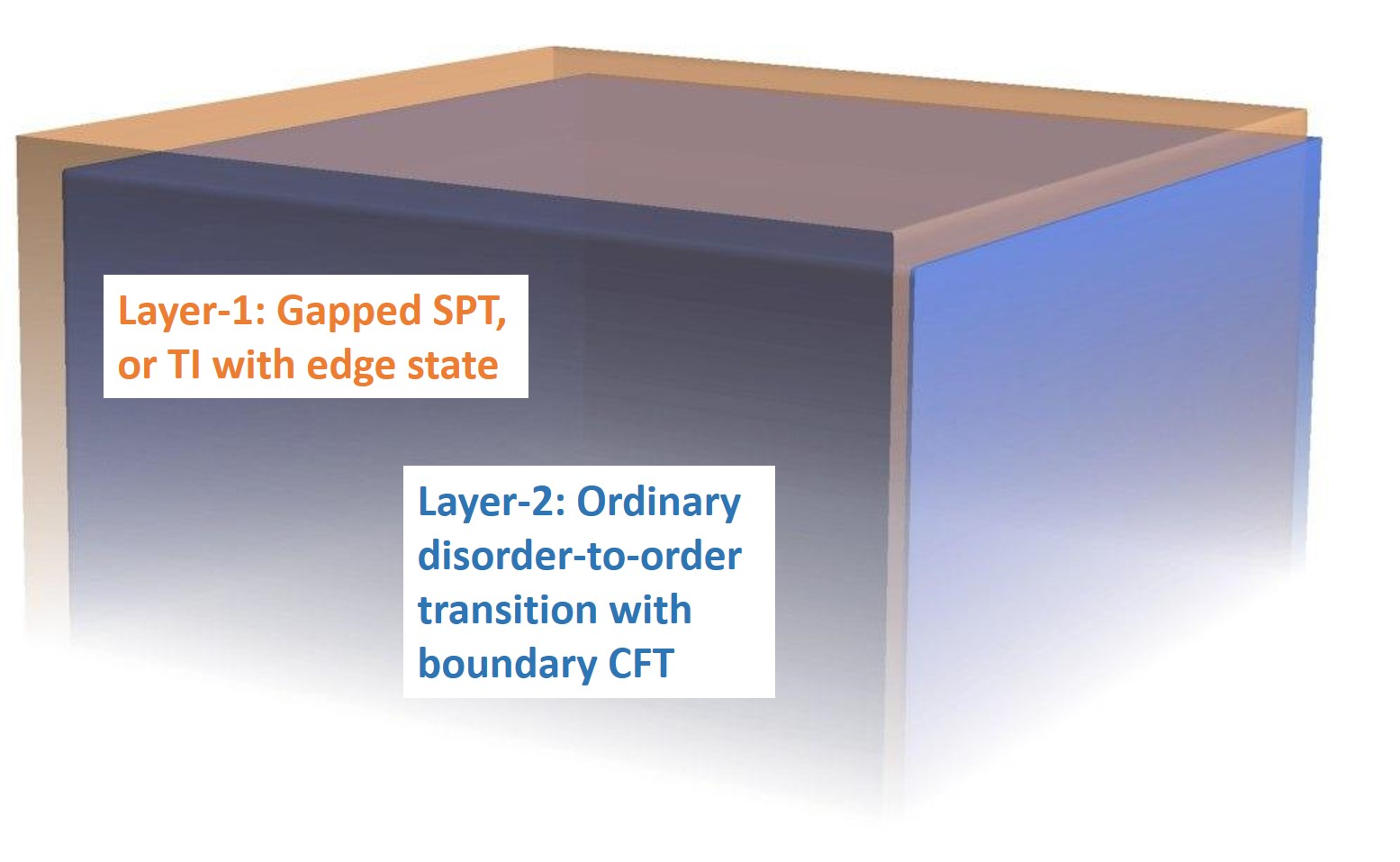}
\caption{We view the system under study as a two layer system.
Layer-1 is a SPT or TI with nontrivial edge states; layer-2 is an
ordinary disorder-to-order phase transition whose order parameter
at the boundary follows the scaling of boundary CFT. The boundary
of the entire system may flow to new fixed points due to the
coupling between the two layers. } \label{layer}
\end{figure}

For our purpose we give the system under study a virtual two-layer
structure Fig.~\ref{layer}: layer-1 is a SPT state with nontrivial
edge states, and it is not tuned to a bulk phase transition;
layer-2 is a topological trivial system which undergoes an
ordinary Landau-Ginzburg disorder-to-order phase transition. Then
as a starting point we assume a weak coupling between the boundary
of the two layers, and study the RG flow of the coupling. Besides
the edge state localized at the boundary of a SPT state, we will
also consider symmetry protected gapless states localized at a
$2d$ interface embedded in a $3d$ bulk. We will demonstrate that
in several cases, including the edge state of a prototype bosonic
SPT state, the $2d$ boundary or interface will flow to a new fixed
point due to the bulk quantum phase transition.

Previous works have explored related ideas with different
approaches. Exactly soluble $1d$ and $2d$ Hamiltonians have been
constructed for gapless systems with protected edge
states~\cite{scaffidi2017}; fate of edge states was also studied
for $1d$ and $2d$ SPT
states~\cite{poll,poll2,zhang,stefan1,stefan2}. But the $2d$ edge
of $3d$ bosonic SPT systems coupled with boundary modes which
originate from bulk quantum critical points, $i.e.$ the situation
that potentially hosts the richest and most exotic phenomena, have
not been studied to our knowledge. We note that the interaction
between bulk quantum critical modes and the boundary of free or
weakly interacting fermion topological insulator (or topological
superconductor) was studied in Ref.~\onlinecite{groveredge}, but
the coupling in that case was strongly irrelevant hence will not
lead to new physics in the infrared (we will review the interplay
between the bulk quantum critical modes and the edge states of
free fermion topological insulator in the next section). We will
focus on bosonic SPT state with intrinsic strong interaction in
this work. We use the generic long wavelength field theory
description of both the bulk bosonic SPT states and the edge
states. Due to the lack of exact results of strongly interacting
$(2+1)d$ field theories, we seek for a controlled calculation
procedure that allows us to identify new fixed points under RG
flow. Indeed, in several scenarios we will explore in this work,
new fixed points are identified based on controlled calculations.

\section{Edge States of $3d$ SPT at Bulk QCP}

\subsection{Edge states of noninteracting $3d$ TIs}

We first consider the edge state of $3d$ topological insulator
(TI) and symmetry protected topological states. The edge state of
free fermion TI is described by the action \beqn \mathcal{S} =
\int d^2x d\tau \sum_{\alpha = 1}^{N_f} \bar{\psi}_\alpha
\gamma_\mu
\partial_\mu \psi_\alpha, \eeqn with $\gamma^1 = \sigma^2$,  $\gamma^2 =
- \sigma^1$, $\gamma^0 = \sigma^3$, $\bar{\psi} = \psi^\dagger
\gamma^0$. Based on the ``ten-fold way
classification"~\cite{ludwigclass1,ludwigclass2,kitaevclass}, for
the AIII class, at the noninteracting level the TI is always
nontrivial and topologically different from each other for
arbitrary integer$-N_f$; while for the AII class the TI is
nontrivial only for odd integer $N_f$, and they are all
topologically equivalent to the simplest case with $N_f = 1$. In
both cases the fermion mass term $\sum_\alpha \bar{\psi}_\alpha
\psi_\alpha$ is forbidden by the time-reversal symmetry. Hence let
us consider the disorder-to-order phase transition in the $3d$
bulk associated with a spontaneous time-reversal symmetry
breaking, which is described by an ordinary $(3+1)d$
Landau-Ginzburg quantum Ising theory: \beqn \mathcal{S}_b = \int
d^3x d\tau \ (\partial \phi)^2 + u \phi^4. \label{ising}\eeqn
Because $u$ is a marginally irrelevant coupling at the $(3+1)d$
noninteracting Gaussian fixed point, the scaling dimension of
$\phi$ in the bulk is precisely $[\phi] = 1$.

Here we stress that the disorder-to-order transition is driven by
the physics in the bulk. Without the bulk, the boundary alone does
not support an ordered phase.
To study the fate of the edge state when the bulk is tuned to the
quantum critical point, we view the bulk as a ``two layer" system
(Fig.~\ref{layer}): layer-1 is a $3d$ TI which is not tuned to the
quantum phase transition; while layer-2 is at the
disorder-to-order bulk quantum phase transition between a
time-reversal invariant trivial insulator and a spontaneous
time-reversal symmetry breaking phase. Now both layers have
nontrivial physics at the edge. The quantum critical fluctuation
(from layer-2) at the $2d$ boundary must satisfy the boundary
scaling law. When we impose the most natural boundary condition
$\phi(z \geq 0) = 0$, the leading field at the boundary which
carries the same quantum number as $\phi$ is $\Phi \sim
\partial_z \phi$. Since $\phi$ has scaling dimension $1$, $\Phi$
should have scaling dimension $[\Phi] = 2$, $i.e.$ \beqn \langle
\Phi(\mathbf{x}, z=0) \Phi(0, z = 0) \rangle \sim
1/|\mathbf{x}|^4, \label{corre} \eeqn where $\mathbf{x} = (\tau,
x, y)$. Eq.~\ref{corre} is a much weaker correlation than $\phi$
in the bulk (more detailed derivation of boundary correlation
functions can be found in
Ref.~\onlinecite{cardybook,boundary2,boundary3,boundary4}).

Now we turn on coupling between the $2d$ boundaries of the two
layers. The edge state of the TI in layer-1 is affected by the
boundary fluctuations of layer-2 through the ``proximity effect".
The coupling between the two layers at the $2d$ boundary is
described by the following term in the action: \beqn \mathcal{S}_c
= \int d^2x d\tau \ \sum_\alpha g \Phi \bar{\psi}_\alpha
\psi_\alpha. \eeqn Since $\Phi \sim \partial_z \phi$ has scaling
dimension $2$, $g$ will have scaling dimension $[g] = -1$, $i.e.$
it is strongly irrelevant. This conclusion is consistent with
previous study Ref.~\onlinecite{groveredge}. A negative ``mass
term" $\Phi^2$ will be generated through the standard fermion loop
diagram, but since $\Phi$ has scaling dimension 2, this mass term
will be irrelevant. Hence the edge state of a $3d$ TI is stable
even at the bulk quantum critical point where the time-reversal
symmetry is spontaneously broken, and the properties of the edge
states (such as electron Green's function) should be identical to
the edge state of TI in the infrared. To make the coupling $g$
relevant, the quantum critical modes also need to localize on the
boundary, which is one of the situations studied in
Ref.~\onlinecite{groveredge}.

\subsection{Edge states of bosonic SPT states}

The situation of bosonic SPT phases can be much more interesting.
The bosonic SPT state can only exist in strongly interacting
systems. We use the prototype $3d$ bosonic SPT phase with $(\U(1)
\times \U(1)) \times Z_2^T$ symmetry as an example, since this
phase can be viewed as the parent state of many $3d$ bosonic SPT
phases by breaking the symmetry down to its subgroups, without
fully trivializing the SPT phase. The topological feature of this
phase can be conveniently captured by the following nonlinear
sigma model in the $(3+1)d$ bulk~\cite{senthilashvin,xuclass}:
\beqn \mathcal{S} = \int d^3x d\tau \ \frac{1}{g} (\partial
\vect{n})^2 + \frac{\ii 2\pi}{\Omega_4} \epsilon_{abcde} n^a
\partial_x n^b \partial_y n^c \partial_z n^d \partial_\tau n^e,
\label{nlsm}\eeqn where $\vect{n}$ is a five component vector
field with unit length, and $\Omega_4$ is the volume of the four
dimension sphere with unit radius. $(n_1, n_2)$, and $(n_3, n_4)$
transform as a vector under the two $\U(1)$ symmetries
respectively, and the $Z_2^T$ changes the sign of all components
of the vector $\vect{n}$. The nonlinear sigma model Eq.~\ref{nlsm}
is invariant under all the transformations.

The $2d$ edge state of this SPT phase can be described by the
following $(2+1)d$ action: \beqn \mathcal{S} &=& \int d^2x d\tau
\sum_{\alpha = 1,2}|(\partial - \ii a) z_\alpha|^2 + r|z_\alpha|^2
+ u |z_\alpha|^4 \cr\cr &+& \frac{1}{e^2} (\mathrm{d} a)^2,
\label{actions} \eeqn where $a_\mu$ is a noncompact $\U(1)$ gauge
field. The theory Eq.~\ref{actions} is referred to as the
``easy-plane noncompact CP$^1$" (EP-NCCP$^1$) model. We are most
interested in the point $r = 0$. The term $\sum_\alpha
r|z_\alpha|^2$ would be forbidden if there is an extra $Z_2$
self-dual symmetry that exchanges the two $\U(1)$
symmetries~\cite{ashvinlesik}, while without the self-duality
symmetry $r$ needs to be tuned to zero, and the point $r = 0$
becomes the transition point between two ordered phases that
spontaneously breaks the two $\U(1)$ symmetries
respectively~\cite{deconfine1,deconfine2}. At $r = 0$, starting
with the UV fixed point with noninteracting $z_\alpha$ and
$a_\mu$, both $u$ and $e$ are expected (though not proven) to flow
to a fixed point with $u = u_\ast$, $e = e_\ast$.

The putative conformal field theory at $r = 0$ and its fate under
coupling to the boundary fluctuations (boundary modes) of the bulk
quantum critical points is the goal of our study in this section.
As was discussed in previous literatures, it is expected that
there is an emergent $\O(4)$ symmetry in Eq.~\ref{actions} at $r =
0$, when we fully explore all the duality features of
Eq.~\ref{actions}~\cite{ashvinlesik,xudual,mrossdual,potterdual,seiberg2,deconfinedual,dualreview}.
In the EP-NCCP$^1$ action, the following operators form a vector
under $\O(4)$: \beqn (n_1, n_2, n_3, n_4) \sim (z^\dagger \sigma^1
z, \ z^\dagger \sigma^2 z, \ \mathrm{Re}[\mathcal{M}_a], \
\mathrm{Im}[\mathcal{M}_a]), \eeqn where $\mathcal{M}_a$ is the
monopole operator (the operator that annihilates a quantized flux
of $a_\mu$). In the equation above, $(n_1, n_2)$ and $(n_3, n_4)$
form vectors under the two $\U(1)$ symmetries respectively. The
emergent $\O(4)$ includes the self-dual $Z_2$ symmetry of the
EP-NCCP$^1$, $i.e.$ the operation that exchanges the two $\U(1)$
symmetries.

Now we consider the $3d$ bulk quantum phase transition between the
SPT phase and the ordered phases that break part of the defining
symmetries of the SPT phase. We first consider two order
parameters: $\phi_0$, $\phi_3$. $\phi_0$ is the order parameter
that corresponds to the self-dual $Z_2$ symmetry; and $\phi_3$ is
a singlet under the emergent $\SO(4)$ but odd under the improper
rotation of the emergent $\O(4)$, and also odd under $Z_2^T$.
Again we view our system as a two layer structure: layer-1 is a
SPT phase with solid edge states described by Eq.~\ref{actions};
layer-2 is a topological-trivial system that undergoes the
transition of condensation of either $\phi_0$ or $\phi_3$. Both
order parameters have an ordinary mean field like transition in
the bulk of layer-2. Again at the boundary, both order parameters
will have very different scalings from the bulk. We assume that
system under study fills the entire semi-infinite space at $z <
0$, then at the boundary plane $z = 0$, the most natural boundary
condition is that $ \phi_0(z \geq 0) = \phi_3(z \geq 0) = 0$,
hence all order parameters near but inside the bulk should be
replaced by the following representations: $\Phi_0 \sim
\partial_z \phi_0$, $\Phi_3 \sim
\partial_z \phi_3$. Both order parameters have scaling dimensions
$2$ at the $(2+1)d$ boundary of layer-2.

Now we couple $\Phi_0$ and $\Phi_3$ to the edge states of layer-1.
The coupling will take the following form: \beqn \mathcal{L}_{c0}
= \sum_{\alpha} g_0 \Phi_0 |z_\alpha|^2, \ \ \mathcal{L}_{c3} =
g_3 \Phi_3 z^\dagger \sigma^3 z. \eeqn The RG flow of coupling
constants $g_{0,3}$ can be systematically evaluated in certain
large$-N$ generalization of the action in Eq.~\ref{actions}: \beqn
\mathcal{S} &=& \int d^2x d\tau \sum_{\alpha = 1,2} \sum_{j =
1}^{N/2} |(\partial - \ii a) z_{j, \alpha}|^2 + u
(\sum_j|z_{j,\alpha}|^2 )^2. \label{actions2}\eeqn The large$-N$
generalization facilitate calculations of the RG flow, but the
down side is that the duality structure and emergent symmetries no
longer exist for $ N > 2$. In the large$-N$ limit of
Eq.~\ref{actions2}, the scaling dimension of the operators under
study is \beqn N \rightarrow + \infty : \ \ [ z^\dagger \sigma^3
z] = [|z|^2] = 2. \eeqn In the equation above, each operator has a
sum of index $j$, which was not written explicitly. Apparently
coupling constants $g_{0,3}$ are both irrelevant with large$-N$
due to the weakened boundary correlation of $\Phi_0$ and $\Phi_3$.

We are seeking for more interesting scenarios when the boundary is
driven to a new fixed point due to the bulk quantum criticality.
For this purpose we consider another order parameter $\vec{\phi}$
which transforms as a vector under one of the two $\U(1)$
symmetries. Here we no longer assume the $Z_2$ self-dual symmetry
on the lattice scale. Again at the boundary $\vec{\phi}$ should be
replaced by $\vec{\Phi} \sim
\partial_z \vec{\phi}$. At the $2d$ boundary, the coupling between
$\vec{\Phi}$ and the edge state of layer-2 reads \beqn
\mathcal{L}_{cv}  = g_{v} \left( \Phi_{1} z^\dagger \sigma^1 z +
\Phi_{2} z^\dagger \sigma^2 z\right). \label{cv}\eeqn In the
large$-N$ limit of Eq.~\ref{actions2}, the scaling dimension of
the operators under study is \beqn N \rightarrow + \infty : \ \
[z^\dagger \sigma^1 z ] = [z^\dagger \sigma^2 z ] = 1. \eeqn Hence
$g_v$ is marginal in the large$-N$ limit, and there is a chance
that $g_v$ could drive the system to a new fixed point with $1/N$
corrections.

We introduce the following action in order to compute the RG flow
of $g_v$ with finite but large $N$: \beqn \mathcal{S} &=& \int
d^2x d\tau \sum_{\alpha = 1,2} \sum_{j = 1}^{N/2} |(\partial - \ii
a) z_{j, \alpha}|^2 + \ii \lambda_+ |z_{j,\alpha}|^2 \cr\cr &+&
\ii \lambda_- z^\dagger_j \sigma^3 z_j + \ii g_v \vec{\Phi} \cdot
z^\dagger_j \vec{\sigma} z_j + \frac{1}{2} \vec{\Phi} \cdot
\frac{1}{|\partial|} \vec{\Phi}. \label{action3} \eeqn The
$\lambda_{\pm}$ are two Hubbard-Stratonovich (HS) fields
introduced for the standard $1/N$
calculations~\cite{kaulsachdev,B2019}. The scaling of $|z|^2$ and
$z^\dagger \sigma^3 z$ in Eq.~\ref{actions2} are replaced by the
HS fields $\lambda_+$, $\lambda_{-}$ in the new action
Eq.~\ref{action3} respectively. A coefficient ``$\ii$" is
introduced in the definition of $g_v$ by redefining $\Phi
\rightarrow \ii \Phi$ for convenience of calculation.

\begin{figure}[h]
\includegraphics[width=0.74\linewidth]{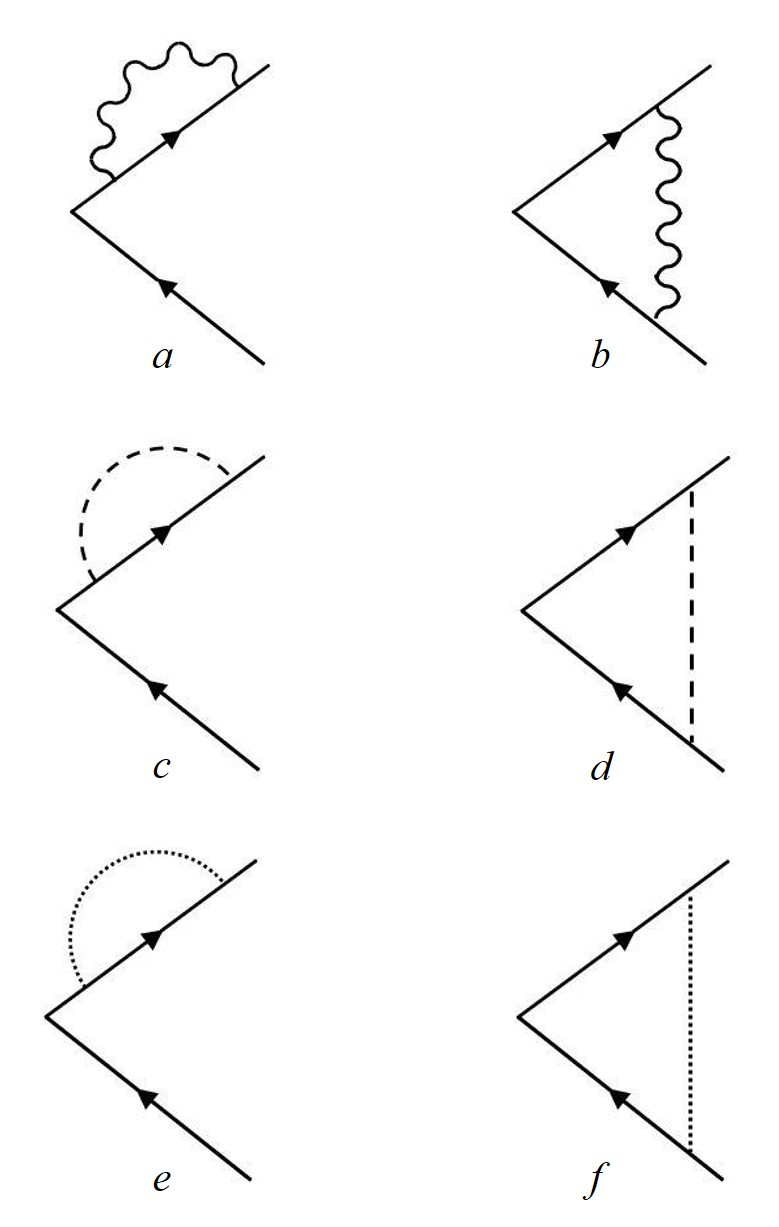}
\caption{$(a,b)$ the $1/N$ contribution to $z^\dagger \sigma^{1,2}
z$ and $\bar{\psi} \tau^{1,2} \psi$ from the gauge field
fluctuation, the solid lines represent either the propagator of
$z_\alpha$ or $\psi_\alpha$, the wavy line represents the
propagator of the photon; $(c,d)$ the $1/N$ contribution to
$z^\dagger \vec{\sigma} z$ from $\lambda_{\pm}$ in
Eq.~\ref{action3}; $(e,f)$ the contribution to $B$ in
Eq.~\ref{beta}. } \label{fd}
\end{figure}

\begin{figure}[h]
\includegraphics[width=0.8\linewidth]{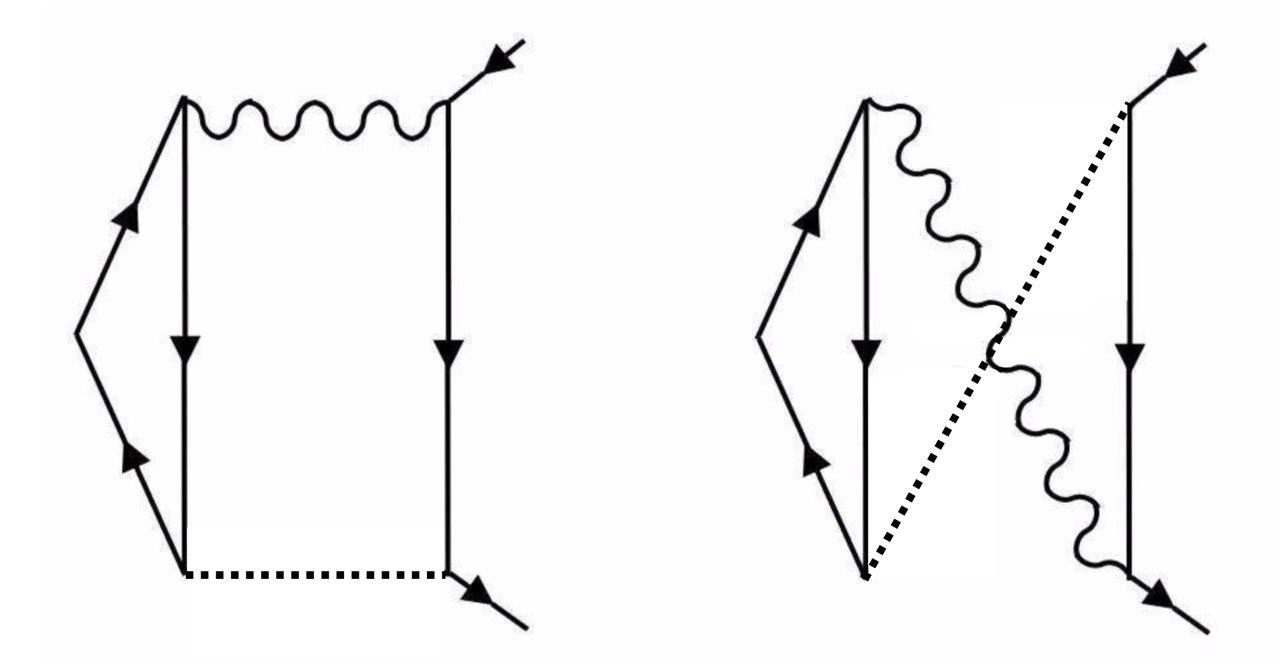}
\caption{The two diagrams at $g_v^3$ order which cancel each other
for arbitrary gauge choices.} \label{fdp}
\end{figure}

The schematic beta function of $g_v$ reads \beqn \frac{dg_v}{d\ln
l} = (1 - \Delta_v) g_v - B g_v^3 + O(v^5). \label{beta}\eeqn
$\Delta_v$ is the scaling dimension of $z^\dagger_j \vec{\sigma}
z_j$ in the large$-N$ generalization of the EP-NCCP$^1$ model
Eq.~\ref{actions2}, with $\vec{\sigma} = (\sigma^1, \sigma^2)$.
The standard $1/N$ calculation leads to \beqn \Delta_v = 1 -
\frac{56}{3\pi^2 N} + O(\frac{1}{N^2}). \label{scalinggv} \eeqn
The $1/N$ correction of $\Delta_v$ comes from diagram
Fig.~\ref{fd}$(a-d)$, where the wavy line is the gauge boson
propagator, and the dashed line represents propagators of both
$\lambda_{\pm}$. The first term of Eq.~\ref{scalinggv} implies
that $g_v$ is indeed weakly relevant with finite but large$-N$.

The constant $B$ in the beta function arises from the operator
product expansion of the coupling term Eq.~\ref{cv}, which is
equivalent to the diagrams Fig.~\ref{fd}$e,f$. This computation
leads to $B = 1/(3\pi^2)$. The two diagrams in Fig.~\ref{fdp}
which are also at $g_v^3$ order cancel each other for arbitrary
gauge choices. Similar two-loop diagrams at the same order of
$1/N$ do not enter the RG equation due to lack of logarithmic
contribution, as was explained in Ref.~\onlinecite{B2019}.
$\vec{\Phi}$ does not receive a wave function renormalization due
to the singular form of its action. Hence with finite but
large$-N$, $g_v$ indeed flows to a new fixed point: \beqn
g_{v\ast}^2 = \frac{56}{N} + O(\frac{1}{N^2}). \eeqn We stress
that this result is drawn from a controlled calculation and it is
valid to the leading order of $1/N$.

\begin{figure}[h]
\includegraphics[width=\linewidth]{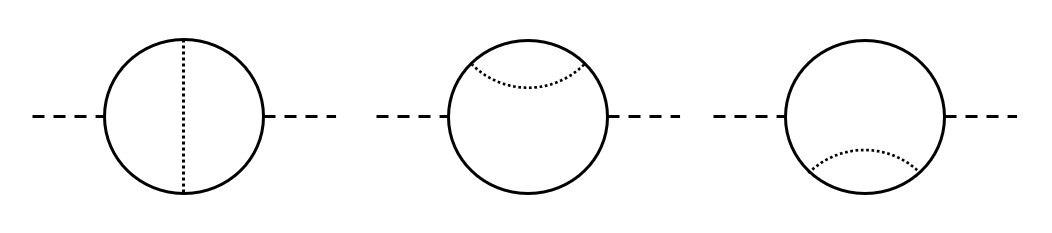}
\caption{The $g_v^2$ diagrams that contributes to the scaling
dimension of $[\lambda_+]$. Here the solid line represents the
propagator of $z_{j,\alpha}$, the dotted line represents the
vector operator $\vec{\Phi}$, and the dashed line represents
$\lambda_+$.} \label{fd5}
\end{figure}

As we explained before, the point $r = 0$ is a direct transition
between two ordered phases that spontaneously break the two
$\U(1)$ symmetries. This transition will be driven to a new fixed
point by coupling to the boundary fluctuations of bulk critical
points as we demonstrated above. At this new fixed point, the
critical exponent $\nu$ follows from the relation \beqn \nu^{-1} =
3- [\lambda_+]. \eeqn To evaluate the scaling dimension
$[\lambda_+]$ we have to incorporate the contributions of $g_v^2$
from the diagrams shown in~\figref{fd5}, and combined with $1/N$
calculations performed previously~\cite{wen-wu,B2019}. Then in the
end we obtain \beqn \nu^{-1}_* &=& 1 + \frac{160}{3\pi^2N} +
\frac{4g_{v*}^2}{3\pi^2} + O(\frac{1}{N^2}) \cr\cr &=& 1 +
\frac{128}{\pi^2N} + O(\frac{1}{N^2}). \eeqn Again, there are
other loop diagrams which appear to be at the same order of $1/N$
but do not make any logarithmic contributions~\cite{B2019}.

\section{Interface States Embedded in $3d$ bulk}

\subsection{Interface states of noninteracting electron systems}

\begin{figure}[h]
\includegraphics[width=0.7\linewidth]{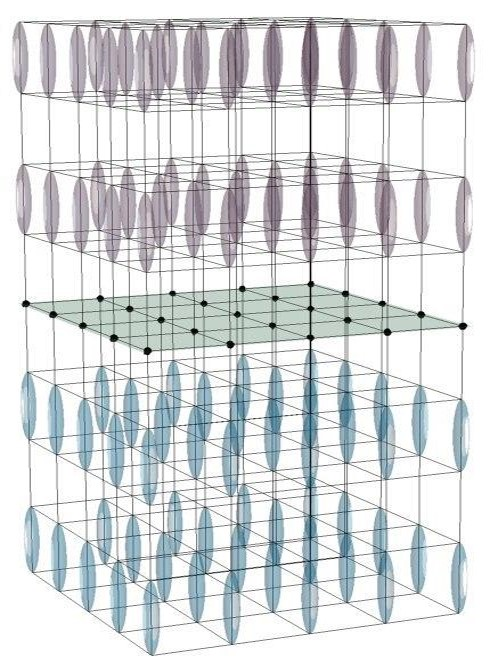}
\caption{We consider a $\SU(N)$ antiferromagnet with
self-conjugate representation on each site. The system forms a
background VBS pattern, with opposite dimerizations between
semi-infinite spaces $z > 0$ and $z < 0$. There is a $2d$
antiferromagnet localized at the interface $z = 0$, and the entire
bulk can undergo phase transition simultaneously due to the mirror
(reflection) symmetry that connects the two sides of the domain
wall.} \label{dimer}
\end{figure}

In previous examples we studied topological edge states at the
boundary of a $3d$ system. In this section we will consider the
$2d$ states localized at an interface ($z = 0$) in a $3d$ space,
when the entire $3d$ bulk (for both $z > 0$ and $z < 0$
semi-infinite spaces) undergoes a phase transition simultaneously.
Without fine-tuning, we need to assume an extra reflection
symmetry $z \rightarrow -z$ that connects the two sides of the
interface, which guarantees a simultaneous phase transition in the
entire system. 
In this case there is no physical reason to impose the strong
boundary condition at the interface embedded in the $3d$ space,
hence the quantum critical modes at the interface follow the
ordinary bulk scalings, instead of the weakened correlation of
boundary CFT.

Again we will consider free fermion systems first. Let us first
recall that the AIII class TI has a $\mathbb{Z}$ classification
which is characterized by a topological index $n_T$. $n_T$ will
appear as the coefficient of the electromagnetic response of the
TI: $\mathcal{L} \sim i \pi n_T \mathbf{E} \cdot \mathbf{B} $.
$n_T$ must change sign under spatial reflection transformation
$\mathcal{M}_z : z \rightarrow -z$. To construct the desired
system, we assume the semi-infinite space $z < 0$ is occupied with
the AIII class TI with Hamiltonian $\hat{H}$, whose topological
index is $n_T$; and its ``reflection conjugate"
$\mathcal{M}_z^{-1} \hat{H} \mathcal{M}_z$ fills the semi-infinite
space $z > 0$. Then there are $N_f = 2n_T$ flavors of massless
Dirac fermions localized at the $2d$ plane $z = 0$, which are
still protected by time-reversal symmetry. Now we assume the
entire bulk undergoes a quantum phase transition with a
spontaneous time-reversal symmetry breaking, whose order parameter
couples to the domain wall Dirac fermions as \beqn \mathcal{S} &=&
\int d^2x d\tau \ \sum_{\alpha = 1}^{N_f} \bar{\psi}_\alpha
\gamma_\mu
\partial_\mu \psi_\alpha + g \phi \bar{\psi}_\alpha \psi_\alpha \cr\cr
&+& \frac{1}{2} \phi (-\partial^2)^{1/2} \phi. \eeqn The last term
in the action is still defined in the $(2+1)d$ interface, and it
reproduces the correlation of $\phi$ in the bulk: $\langle
\phi(0)\phi(r) \rangle\sim 1/r^2 $. We stress that, since now the
order parameter resides in the entire bulk, $\phi$ no longer obeys
the boundary scaling as we discussed in previous examples. A
negative boson mass term $- r \phi^2$ can be generated through the
standard fermion mass loop diagram, hence we need to tune an extra
term at the interface to make sure the mass term of $\phi$
vanishes.

In this case the coupling constant $g$ is a marginal perturbation
based on simple power-counting. But $g$ will flow under
renormalization group (RG) with loop corrections in
Fig.~\ref{fd}$(e,f)$: \beqn \beta(g) = \frac{dg}{d\ln l} = -
\frac{2}{3\pi^2}g^3 + O(g^5). \label{beta1}\eeqn Hence even in
this case, the coupling between the domain wall states and the
bulk quantum critical modes is perturbatively marginally
irrelevant.

\begin{figure}[h]
\includegraphics[width=1\linewidth]{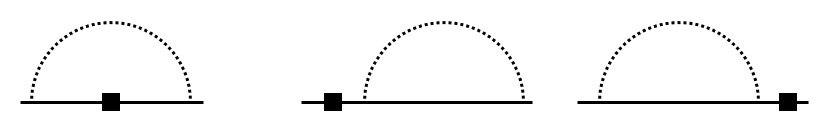}
\caption{The Feynman diagrams that renormalizes the extra velocity
$\delta$ in Eq.~\ref{delta}. The box represents the vertex
$\delta$, and all three diagrams contributes to the fermion
self-energy and renormalize $\delta$. } \label{fd6}
\end{figure}

So far we have assumed that the velocity of the interface state is
identical with the bulk. Now let us tune the velocity of the
domain wall Dirac fermions slightly different, which can be
captured by the following term in the Lagrangian: \beqn
\sum_\alpha \delta \bar{\psi}_\alpha (\gamma^1
\partial_x + \gamma^2 \partial_y - 2 \gamma^3 \partial_3)\psi_\alpha.
\label{delta}\eeqn $\delta$ defined above is an eigenvector under
the leading order RG flow. With the loop diagrams in \figref{fd6},
we obtain the leading order beta function of $\delta$: \beqn
\beta(\delta) =\frac{d\delta}{d \ln l}=-\frac{1}{5\pi^2}g^2\delta.
\eeqn Together with $\beta(g)$, the velocity anisotropy is also
perturbatively irrelevant.

\begin{figure}[h]
\includegraphics[width=0.75\linewidth]{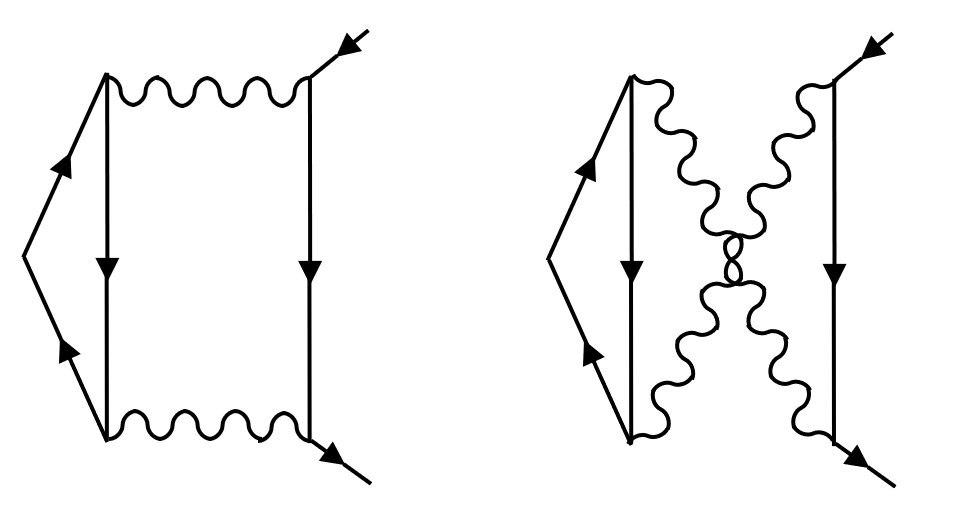}
\caption{The extra diagrams that contribute to the scaling
dimension of $\sum_\alpha \bar{\psi}_\alpha \psi_\alpha$ at the
leading order of $1/N_f$ in QED$_3$. Again the wavy lines are
photon propagators. } \label{fd4}
\end{figure}

\subsection{Interface states of quantum antiferromagnet}

We now consider a $\SU(N)$ quantum antiferromagnet on a tetragonal
lattice with a self-conjugate representation on each site (we
assume $N$ is an even integer). With large$-N$, an
antiferromagentic Heisenberg $\SU(N)$ model has a dimerized ground
state~\cite{rokhsar,sachdevread} where the two $\SU(N)$ spins on
two nearest neighbor sites form a spin singlet (valence bond). We
consider the following background configuration of valence bond
solid (VBS): the spins form VBS along the $\hat{z}$ direction
which spontaneously break the translation symmetry, while there is
a domain wall between two opposite dimerizations at the $2d$ XY
plane $z = 0$, namely $z = 0$ is still a mirror plane of the
system (Fig.~\ref{dimer}). In each $1d$ chain along the $\hat{z}$
direction, there is a dangling self-conjugate $\SU(N)$ spin
localized on the site at the domain wall. Hence the $2d$ domain
wall is effectively a $\SU(N)$ antiferromagnet on a square
lattice.

One state of $\SU(N)$ antiferromagnet which is the ``parent" state
of many orders and topological orders on the square lattice, is
the gapless $\pi-$flux $\U(1)$ spin
liquid~\cite{ianbrad,hermele2005}. At low energy this spin liquid
is described by the following action of $(2+1)d$ quantum
electrodynamics (QED$_3$): \beqn \mathcal{S} = \int d^2x d\tau \
\sum_{\alpha = 1}^{N_f} \bar{\psi}_\alpha \gamma_\mu (\partial_\mu
- \ii a_\mu) \psi_\alpha + \cdots \eeqn $\psi_\alpha$ is $N_f =
2N$ flavors of $2-$component Dirac fermions, and they are the low
energy Dirac fermion modes of the slave fermion $f_{j,\alpha}$
defined as $\hat{S}^b_j = f^\dagger_{j,\alpha} T^b_{\alpha\beta}
f_{j,\beta} $, $T^b$ with $b = 1 \cdots N^2 - 1$ are the
fundamental representation of the $\SU(N)$ Lie Algebra. Besides
the spin components, there is an extra two dimensional internal
space which corresponds to two Dirac points in the Brillouin zone.
There is an emergent $\SU(N_f)$ flavor symmetry in QED$_3$ which
includes both the $\SU(N)$ spin symmetry and discrete lattice
symmetry.

It is known that when $N_f$ is greater than a critical integer,
the QED$_3$ is a conformal field theory (CFT). We will consider
the fate of this CFT when the three dimensional bulk is driven to
a quantum phase transition. We will first consider a
disorder-to-order quantum phase transition, where the ordered
phase spontaneously breaks the time-reversal and parity symmetry
of the XY plane. Notice that due to the reflection symmetry $z
\rightarrow -z$ of the background VBS configuration, the two sides
of the domain wall will reach the quantum critical point
simultaneously. The bulk transition is still described by
Eq.~\ref{ising}. When we couple the Ising order parameter $\phi$
to the domain wall QED$_3$, the total $(2+1)d$ action reads \beqn
\mathcal{S} &=& \int d^2x d\tau \ \sum_{\alpha = 1}^{N_f}
\bar{\psi}_\alpha \gamma_\mu (\partial_\mu - \ii a_\mu)
\psi_\alpha \cr\cr &+& g \phi \bar{\psi}_\alpha \psi_\alpha +
\frac{1}{2} \phi (-\partial^2)^{1/2} \phi. \eeqn If the gauge
field fluctuation is ignored, or equivalently in the large$-N_f$
limit, the scaling dimension of $\bar{\psi}\psi$ is
$[\bar{\psi}\psi] = 2$, and hence the scaling dimension of $g$ is
$[g] = 0$, $i.e.$ $g$ is a marginal perturbation. The $1/N_f$
correction to the RG flow arises from the Feynman diagrams
(Fig.~\ref{fd}$(a,b)$ and Fig.~\ref{fd4}) which involves one or
two photon propagators: \beqn G^a_{\mu\nu}(\vec{p}) =
\frac{16}{N_f p} \left( \delta_{\mu\nu} - \frac{p_\mu p_\nu}{p^2}
\right).  \eeqn Again in this case the fermions will generate a
mass term for the order parameter at the interface, which we need
to tune to zero. At the leading order of $1/N_f$ the corrected
beta function for $g$ reads \beqn \beta(g) = \frac{dg}{d\ln l} = -
\frac{128}{3\pi^2 N_f} g - \frac{2}{3\pi^2} g^3 + O(g^3).
\label{beta1} \eeqn But this beta function does not lead to a new
unitary fixed point other than the decoupled fixed point $g = 0$.
Hence in this case the domain wall state is decoupled from the
bulk quantum critical modes in the infrared limit.

A more interesting scenario is when the bulk undergoes a
transition which spontaneously breaks the translation and $C_4$
rotation symmetry by developing an extra VBS order within the XY
plane. The inplane VBS order parameters are $V_x \sim \bar{\psi}
\tau^1 \psi$, and $V_y \sim \bar{\psi} \tau^2 \psi$, where
$\tau^{1,2}$ are the Pauli matrices operating in the Dirac valley
space. The coupling between the VBS order parameter and the domain
wall QED$_3$ reads \beqn \mathcal{S}_c = \int d^2x d\tau \ g
\left( \phi^\ast \bar{\psi} \tau^- \psi + \phi \bar{\psi} \tau^+
\psi \right) + \phi^\ast (- \partial^2)^{1/2} \phi. \eeqn Here
$\tau^{\pm } = (\tau^1 \pm \ii \tau^2)/2$. The scaling dimension
of the VBS order parameter at the QED$_3$ fixed point has been
computed previously~\cite{hermele2005,ranwen,xusachdev}:
$[\bar{\psi}\tau^a \psi] = 2 - 64 / (3 \pi^2 N_f)$, and the beta
function of $g$ to the leading order of $1/N_f$ reads \beqn
\beta(g) = \frac{64}{3\pi^2 N_f} g - \frac{1}{6\pi^2 } g^3 +
O(g^3). \label{beta2} \eeqn In the large$-N_f$ limit, the coupling
$g$ is marginally irrelevant; but with finite and large$-N_f$, $g$
is weakly relevant at the noninteracting fixed point, and it will
flow to an interacting fixed point \beqn g_\ast^2 =
\frac{128}{N_f} + O(\frac{1}{N_f^2}). \eeqn

This new fixed point will break the emergent $\SU(N_f)$ flavor
symmetry down to $\SU(N) \times U(1)$ symmetry, where $\U(1)$
corresponds to the rotation of the Dirac valley space. The
following gauge invariant operators receive different corrections
to their scaling dimensions from coupling to the bulk quantum
critical modes: \beqn [\bar{\psi}\psi] &=& 2 + \frac{128}{3\pi^2
N_f} + \frac{2}{3\pi^2}g_\ast^2 + O(\frac{1}{N_f^2}); \cr\cr
[\bar{\psi}T^b \psi] &=& 2 - \frac{64}{3\pi^2 N_f} +
\frac{2}{3\pi^2} g_\ast^2 + O(\frac{1}{N_f^2}); \cr\cr [\bar{\psi}
\tau^3 \psi] &=& 2 - \frac{64}{3\pi^2 N_f} -
\frac{1}{3\pi^2}g_\ast^2 + O(\frac{1}{N_f^2}); \cr\cr
[\bar{\psi}\tau^{1,2}\psi] &=& 2 - \frac{64}{3\pi^2 N_f} +
\frac{1}{6\pi^2}g_\ast^2 . \label{operator}\eeqn The operators
$\bar{\psi}\tau^{1,2}\psi$ have exactly scaling dimension 2, the
Feynman diagram contributions from Fig.~\ref{fd} cancel each other
for operator $\bar{\psi}\tau^{1,2}\psi$ as they should. Notice
that the last three operators in Eq.~\ref{operator} should have
the same scaling dimension in the original QED$_3$ fixed point due
to the large $\SU(N_f)$ flavor symmetry, but at this new fixed
point they will acquire different corrections.

Another interesting scenario is that the bulk is at a critical
point whose order parameter couples to the Ising like operator
$\bar{\psi}\tau^3 \psi$, which breaks the inplane parity but
preserves the time-reversal: \beqn \mathcal{S}_c = \int d^2x d\tau
\ g \phi \bar{\psi} \tau^3 \psi + \frac{1}{2} \phi (-
\partial^2)^{1/2} \phi. \label{action4} \eeqn The microsopic representation
of the operator $\bar{\psi}\tau^3 \psi$ can be found in
Ref.~\onlinecite{hermele2005}. The beta function of the coupling
$g$ reads \beqn \beta(g) = \frac{64}{3\pi^2 N_f} g -
\frac{2}{3\pi^2 } g^3 + O(g^3), \label{beta3} \eeqn and once again
there is new stable fixed point $g_\ast^2 = 32/N_f + O(1/N_f^2)$.
And at this fixed point, \beqn [\bar{\psi}\psi] &=& 2 +
\frac{128}{3\pi^2 N_f} + \frac{2}{3\pi^2}g_\ast^2 +
O(\frac{1}{N_f^2}); \cr\cr [\bar{\psi}T^b \psi] &=& 2 -
\frac{64}{3\pi^2 N_f} + \frac{2}{3\pi^2} g_\ast^2 +
O(\frac{1}{N_f^2}); \cr\cr [\bar{\psi}\tau^{1,2}\psi] &=& 2 -
\frac{64}{3\pi^2 N_f} - \frac{1}{3\pi^2}g_\ast^2 +
O(\frac{1}{N_f^2}); \cr\cr [\bar{\psi} \tau^3 \psi] &=& 2 -
\frac{64}{3\pi^2 N_f} + \frac{2}{3\pi^2}g_\ast^2.
\label{operator2} \eeqn

The domain wall state considered here is formally equivalent to
the boundary state of a $3d$ bosonic SPT state with pSU$(N) \times
\U(1)$ symmetry, which can also be embedded to the $3d$ SPT with
pSU($N_f$) symmetry discussed in Ref.~\onlinecite{xu3dspt}. This
SPT state can be constructed as follows: we first break the
$\U(1)$ symmetry in the $3d$ bulk by driving the bulk $z < 0$ into
a superfluid phase, and then decorate the vortex loop of the
superfluid phase with a $1d$ Haldane phase with pSU$(N)$
symmetry~\cite{psuhaldane1,psuhaldane2,psuhaldane3,psun}.
Eventually we proliferate the decorated vortex loops to restore
all the symmetries in the bulk. A $1d$ pSU$(N)$ Haldane phase can
be constructed as a spin-chain with a pSU$(N)$ spin on each site,
and there is a dangling self-conjugate representation of $\SU(N)$
on each end of the chain. And this dangling spin will also exist
in the $\U(1)$ vortex at the boundary of the pSU$(N) \times \U(1)$
SPT state. Notice that the self-conjugate representation of
$\SU(N)$ is a projective representation of pSU($N$).

\section{Discussion}

In this work we systematically studied the interplay of two
different nontrivial boundary effects: the $2d$ edge states of
$3d$ symmetry protected topological states, and the boundary
fluctuations of $3d$ bulk quantum phase transitions. New fixed
points were identified through generic field theory descriptions
of these systems and controlled calculations. We then generalized
our study to the $2d$ states localized at the interface embedded
in the $3d$ bulk.

The last case studied in Eq.~\ref{beta3}, \ref{operator2} is
special when $N_f = 2$, and when the gauge field is noncompact.
This is the theory that has been shown to be dual to the
EP-NCCP$^1$ model~\cite{potterdual,deconfinedual} studied in
Eq.~\ref{actions}, the operator $ \sum_\alpha r |z_\alpha|^2$ is
dual to $r \bar{\psi} \tau^3 \psi$, and both theories are
self-dual. By coupling the operator $\bar{\psi}\tau^3\psi$ to the
bulk critical modes (rather than the boundary fluctuations of the
bulk critical points), we have shown that this $(2+1)d$ theory is
driven to a new fixed point, and the self-duality structure still
holds. The self-duality transformation of Eq.~\ref{actions} now is
combined with the Ising symmetry of the order parameter $\phi$.
However, the $\O(4)$ emergent symmetry no longer exists at this
new fixed point, due to the nonzero fixed point of $g$ in
Eq.~\ref{action4}.

The methodology used in this work can have many potential
extensions. We can apply the same field theory and RG calculation
to the $1d$ boundary of $2d$ SPT states (for instance the AKLT
state), which was studied through exactly soluble lattice
Hamiltonians~\cite{scaffidi2017} and also numerical
methods~\cite{zhang,stefan1,stefan2}. Also, $1d$ defect in a $3d$
topological state can also have gapless
modes~\cite{randefect,teodefect}, it would be interesting to
investigate the fate of a $1d$ defect embedded in a $3d$ bulk at
the bulk quantum phase transition. Defects of free or weakly
interacting fermionic topological insulator and topological
superconductor coupled with bulk critical modes was studied in
Ref.~\onlinecite{groveredge}, but we expect the defect of an
intrinsic strongly interacting topological state can lead to much
richer physics. Last but not least, the ``higher order topological
insulator" has nontrivial modes localized at the corner instead of
the boundary of the system~\cite{highorder}. The coupling between
the bulk quantum critical points and corner topological modes is
also worth exploration.

This work is supported by NSF Grant No. DMR-1920434, the David and
Lucile Packard Foundation, and the Simons Foundation. The authors
thank Andreas Ludwig and Leon Balents for helpful discussions.

\bibliography{sb}

\end{document}